# Tailoring light holes in β-Ga$_2$O$_3$ via Anion-Anion Antibonding Coupling

Ke Xu[1†], Qiaolin Yang[2,3†], Wenhao Liu[2], Rong Zhang[1], Zhi Wang[2*], and Jiandong Ye[1*]

[1] School of Electronic Science and Engineering, Nanjing University, Nanjing 210023, China.

[2] State Key Laboratory of Superlattices and Microstructures, Institute of Semiconductors, Chinese Academy of Sciences, Beijing 100083, China

[3] School of Physics and Zhejiang Province Key Laboratory of Quantum Technology and Device, Zhejiang University, Hangzhou 310027, China.

[†] K. X. and Q.-L. Y. contributed equally to this work.

**AUTHOR CONTRIBUTIONS**

K.X. and Q.-L.Y. contributed equally to this work. J.-D.Y., R.Z. and Z.W. conceived and supervised the project. K.X. and Q.-L.Y. developed the idea, designed the research, and implemented the computational algorithm. K.X. and Q.-L.Y. collated the computational data, performed the electronic structure calculations. K.X., Q.-L.Y., W.-H.L., Z.W., and J.D.Y. analyzed the AAAC mechanism and polaronic effect, plotted figures, and wrote the manuscript.

A significant limitation of wide-bandgap materials is their low hole mobility related to localized holes with heavy effective masses ($m_h^*$). We identify in low-symmetric wide-bandgap compounds an anion-anion antibonding coupling (AAAC) effect as the intrinsic factor behind hole localization, which explains the extremely heavy $m_h^*$ and self-trapped hole (STH) formation observed in gallium oxide (β-Ga$_2$O$_3$). We propose a design principle for achieving light holes by manipulating AAAC, demonstrating that specific strain conditions can reduce $m_h^*$ in β-Ga$_2$O$_3$ along c$^*$ from 4.77 $m_0$ to 0.38 $m_0$, making it comparable to the electron mass (0.28 $m_0$), while also slightly suppresses the formation of self-trapped holes, evidenced by the reduction in the formation energy of hole polarons from -0.57 eV to -0.45 eV under tensile strain. The light holes show significant anisotropy, potentially enabling two-dimensional transport in bulk material. This study provides a fundamental understanding of hole mass enhancement and STH formation in novel wide-bandgap materials and suggest new pathways for engineering hole mobilities.

# I. INTRODUCTION

A well-known challenge in the development of wide-bandgap material devices lies in how to reduce the difference in magnitude between electron and hole mobilities. Compounds such as SiC, ZnO, and $Ga_2O_3$ [1–3] normally exhibit relatively larger electrons mobility than those of holes, as well as hold strong hole-phonon interactions that can cause significant carrier scattering, leading to reduced carrier lifetimes [2]. The phenomenon of self-trapped holes (STH), or hole polarons, where excess holes are trapped by the local potential wells created by themselves distorting the lattice nearby, provides evidence of this strong hole-phonon coupling [4–6]. These effects constrain the delocalization of hole states and inherently limit the *intrinsic* hole mobility in wide-gap materials. To enhance hole mobility, it is crucial to investigate the *physical origins* of the heavy $m_h^*$ and the formation of STH.

Although significant progress has been made in synthesizing high-quality, low-cost single crystals and heterojunctions of wide-gap compounds [7–9], the understanding of hole mass enhancement and hole polaron formation remains incomplete and contentious. In a conventional view, the observed heavy $m_h^*$ is attributed to the orbital composition of the VBM. For example, in SiC, GaN, and ZnO, the VBMs predominantly comprise the $2p$ (or, $2s$-$2p$ hybrid) orbitals of C, N, and O, respectively [10–12]. These orbitals have low principal quantum numbers, resulting in deep energy levels and strong localization. Consequently, it appears that the hole masses in these compounds are inherently difficult to tune, given that the electronegativity of the anion and the orbital quantum number are fundamental properties.

However, recent discussions suggest that even in the pristine, defect-free crystals, the orbital composition of the band edge may not exclusively account for the heavy $m_h^*$. Evidence includes studies demonstrating that the hole mass can be remarkably lightened by strain, e.g., in GaN [13], which was explained as the band order reversion caused by changes in crystal-field splitting. Nonetheless, this presents difficulties in explaining the magnitude difference in hole mass between wide-gap materials with the same orbital composition of VBM. For instance, $Ga_2O_3$, a wide-gap compound with an ultraviolet bandgap and a complex phase diagram, has garnered substantial attention [14–16]. Its hole mass has been reported as unusually heavy in both photoemission experiments [17–19] and *ab initio* calculations [20–24], holding a range from 3 $m_0$ up to 40 $m_0$ ($m_0$ is the bare electron mass). This contrasts with materials like SiC, GaN, and ZnO, which all have VBMs composed of anion $2p$ orbitals, yet exhibit more moderate hole effective masses of ~0.6-1.0 $m_0$ [25,26], 1.1~1.6 $m_0$ [27,28], and ~0.8 $m_0$ [3], respectively Notably, the precise evaluation of hole effective mass of Ga$_2$O$_3$ remains challenging through direct measurements, angle-resolved photoemission spectroscopy (ARPES) results has shown a very flat VBM in Ga$_2$O$_3$ [19], which is consistent with theoretical calculations. Therefore, without modifying the valence band of Ga$_2$O$_3$, it is challenging to achieve light holes with enhanced mobility in gallium oxide. Moreover, spontaneous self-trapping of holes (STH) has been observed in high-quality properties [6,14,29]. All these findings suggest that there may be additional, yet overlooked or neglected, factors that dominate the hole mass and the formation of STH.

Some researchers have attempted p-type doping in Ga$_2$O$_3$ by both theoretical prediction and experimental demonstrations, although the stability and reliability of p-type conduction are still questionable and has not been fully verified in bipolar devices [30–33]. Assuming these proposed methods are effective, their common physical mechanism is inducing light holes in Ga$_2$O$_3$ through valence band hybridization. However, an emerging concern is that, heavy doping or alloying may introduce new hybridized bands within the bandgap, potentially reducing the bandgap of Ga$_2$O$_3$, which, in turn, decreases the critical breakdown electric field, making it less favorable for applications in power devices. In addition, the hybridization of the VBM can lead to the formation of new impurity bands due to the extension of impurity energy levels. As a result, holes in the valence band are frequently captured by acceptors and then re-released into the valence band, which significantly reduces hole mobility. Nevertheless, some studies on p-type Ga$_2$O$_3$ remain highly controversial, primarily due to issues such as polaron hopping conduction [34] and grain boundary induced transport in polycrystalline Ga$_2$O$_3$ [31]. To achieve p-type conductivity in Ga$_2$O$_3$, the following conditions must be simultaneously satisfied: a) shallow acceptor dopant levels; b) high free hole density (suppression of STH formation); and c) low hole effective mass. A more detailed discussion is presented in Section VIII.

In this work, we aim to address the gap between experimental observations and physical understandings. Using density functional theory (DFT) for the ground-state β-Ga$_2$O$_3$, we achieved excellent agreements with experimental observations for both the lattice and electronic structures (Table I). Our results reveal that the dominant mechanism of the hole mass enhancement and STH formation is an anion-anion antibonding coupling (AAAC) between several oxygen pairs. Despite the distances between these oxygens being significantly larger than the covalent radius, the strengths of AAAC are remarkably high. This insight led us to the design principle for tuning the hole masses and STH by modifying the strength of AAAC, following which we investigated the effects of strain on the hole states, as strain is a straightforward method to control inter-atom distances hence the coupling. We found that uniaxial tensile strain along the $b$ axis or biaxial compressive strain along $a$-$c$ plane can modulate the AAAC effect as expected and achieve step-function changes on the hole masses. Specifically, the $m^*_{h\|c^*}$ (hole mass along $c^*$) decreases to 8% of its original value, from 4.77 $m_0$ to 0.38 $m_0$, making it comparable to the mass of the electron (0.28 $m_0$). The conductivity mass $m^*_h$, $(m^*_h)^{-1} = \left[\left(m^*_{h\|a^*}\right)^{-1} + \left(m^*_{h\|b^*}\right)^{-1} + \left(m^*_{h\|c^*}\right)^{-1}\right]/3$, also decreases to 29% of its original state, from 3.47 $m_0$ to 0.99 $m_0$. The critical strains for such a transition are 1.5% for the uniaxial tensile strain and 0.7% for the biaxial compressive strain. The resulting light holes are highly anisotropic with $m^*_{h\|b*} \approx 70 m^*_{h\|c*}$, indicating potential for low-dimensional transport in pristine bulk Ga₂O₃ without the need for interfaces. Strain also slightly enhances the density of mobile holes as evidenced by the decreased formation energy of hole polaron. Strain engineering of the valence band in β-Ga₂O₃ offers a promising pathway to enhance p-type conductivity, but further studies are needed to tackle the challenges of deep acceptor levels and stable STH to fully realize p-type conductivity in β-Ga₂O₃. We believe that understanding and manipulating the AAAC effect to control hole mass and STH could pave the way for the design of next-generation wide-gap materials with tailored electronic properties.

## II. THE CRYSTALLOGRAPHIC AND ELECTRONIC STRUCTURES OF β-Ga₂O₃.

Fig. 1 shows the (a) conventional (20-atom) and (b) primitive (10-atom) crystallographic structures of the β-Ga₂O₃. We adopt the lattice vectors and internal atomic coordinates fully relaxed by DFT. The calculated electronic band structure in the first BZ is plotted in Fig. 1(c), while the first BZ and k-points are presented in the inset. All our DFT calculations use the HSE exchange-correlation functional [35,36]. The computational details are given in the Appendix section. The calculated band structure holds an indirect, 4.91 eV gap between the CBM on $\Gamma$ and the global VBM on $M_2$-$D$ path (marked as point $I$) 32 meV higher than the VBM on $\Gamma$ (A more detailed band structure can be found in the supporting information Fig. S1). As the validation of our results, we summarize in Table I the comparisons between this work and experimental observations. All lattice and electronic properties are in good agreement. Note that there are arguments about the position of global VBM. The experimental observation of accurate VBM is yet uncertain due to the wide gap and the flat band edge; in experiment it is reported to be at or nearby $M$ (identical to $M_2$) and about 50 meV higher than that on $\Gamma$ [17–19,37,38] while in previous theoretical works it is claimed to be on the $M_2$-$D$ path [39]. Our calculated hole masses along the three reciprocal directions $a^*$, $c^*$, and $c^*$ (directions are shown in the inset of Fig. 1(c)) are 3.06, 3.06, and 4.77 $m_0$, respectively, while the conductivity mass $m_h^*$ is 3.47 $m_0$. Note that these masses are significantly larger than that of heavy holes in SiC (0.6~1.0 $m_0$) [25,26], GaN (1.1~1.6 $m_0$) [27,28], and ZnO (~0.8 $m_0$) [3].

**Table I**: Comparison of properties in monoclinic β-Ga$_2$O$_3$ from experiments, previous theoretical works, and from this work. Hole masses are calculated along three reciprocal directions $a^*$, $b^*$ and $c^*$, while the electron mass is isotropic. All our DFT results use the HSE exchange-correlation functional [35,36].

| | | Experiment | Theory | This work |
|---|---|---|---|---|
| **Crystallographic** | $a$ (Å) | 12.23 ± 0.02 [a] | 12.25 [b] | 12.25 |
| | $b$ (Å) | 3.04 ± 0.01 [a] | 3.05 [b] | 3.04 |
| | $c$ (Å) | 5.80 ± 0.01 [a] | 5.84 [b] | 5.80 |
| | $\beta$ (°) | 103.7 ± 0.3 [a] | 103.9 [b] | 103.83 |
| **Electronic** | Gap (eV) | 4.85 (indirect) [c] | 4.83 (indirect) [b] | 4.91 (indirect) |
| | Electron mass ($m_0$) | 0.28 ± 0.01 [c] | 0.28 [b] | 0.28 |
| | Hole mass ($m_0$) | 18.75 [d] | 40 ($b^*$), 0.4 ($c^*$) [b]<br>6.14 ($xx$), 2.9 ($yy$), 4.19 ($zz$) [e]<br>3.3 [f], 8.72 [g] | 3.06 ($a^*$), 3.06 ($b^*$), 4.77 ($c^*$) |
| | CBM in k-space | $\Gamma$ [c] | $\Gamma$ [b] | $\Gamma$ |
| | VBM in k-space | $M$ [c] [*] | along $M-A$ [b][e] [*]<br>along $M_2-D$ [h] | along $M_2-D$ |

[a] ref [40] ; [b] ref [20] ; [c] ref [17] ; [d] ref [41] ; [e] ref [22] ; [f] ref [21] ; [g] ref [23] ; [h] ref [39] ; [*] the points $M$ (0.5, 0.5, 0.5) and $M_2$ (-0.5, 0.5, 0.5) are identical; the paths $M-A$ and $M_2-D$ are the same direction along $b^*$.

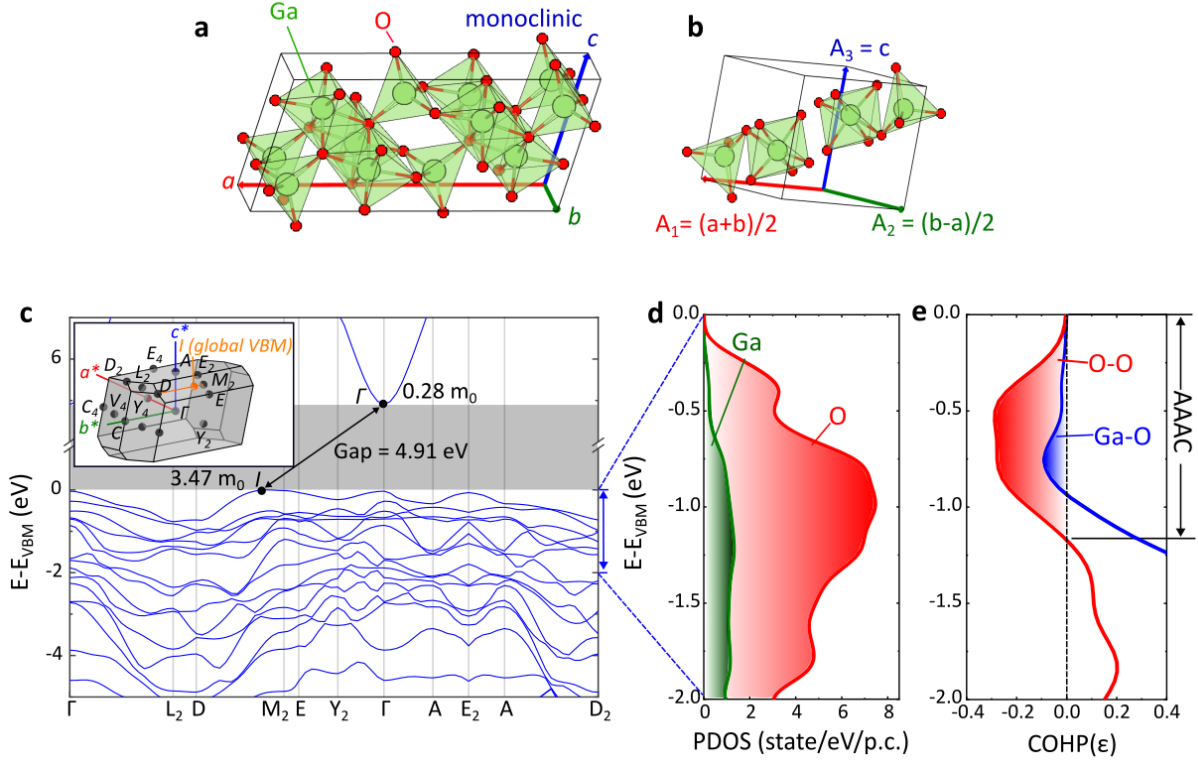

**FIG. 1. Crystal structure, band structure, and atomic orbital coupling of β-Ga₂O₃.** (a) The conventional and (b) the primitive cell, where the shaded green areas are (GaO₆) and (GaO₄) polyhedrons. (c) The electronic band structure calculated by DFT, showing the indirect gap between $\Gamma$ (CBM) and $I$ (global VBM, marked by orange) on $M_2$-$D$ path, and the calculated conductivity effective masses. Inset of (c) shows the primitive first Brillouin zone. Note that the orientations in (b) and inset of (c) are consistent. (d) The atomic orbital projected density of states near VBM onto Ga and O. (e) The projected density of COHP near VBM over all O-O pairs (red) and over all Ga-O pairs (blue), where positive and negative values indicate bonding and antibonding coupling. We mark in (e) the energy range where the O-O antibonding coupling dominates. All our DFT results use the HSE exchange-correlation functional.

### III. THE OXYGEN-DOMINATED VBM AND THE ANION-ANION ANTIBONDING COUPLING.

To understand the physical origin of the heavy hole mass, we firstly do the projection of the density of states (PDOS) onto different atomic orbitals. As shown in Fig. 1(d), within an energy range from -2 eV to 0 eV (VBM has been chosen as energy zero), the DOS comes mostly from oxygen orbitals with only negligible contribution from gallium, indicating an anion-dominated VBM. We then investigate the coupling between different orbitals by calculating the crystal orbital Hamiltonian population (COHP) [42]. The absolute value of COHP represents the coupling strength between specific orbitals, while its sign reveals either such coupling is bonding (if the sign is positive) or antibonding (if it is negative). Details about the COHP calculations are given in the Appendix section. In this work, we mark the COHP between the $\mu$ orbital of the $i^{th}$ atom and the $\nu$ orbital of the $j^{th}$ atom at a given energy $\varepsilon$ as $\text{COHP}_{\mu i, \nu j}(\varepsilon)$.

We firstly calculate the total COHP for all Ga-O coupling and all O-O coupling by using

$$\text{COHP}_{\text{Ga}-\text{O}}(\varepsilon) = \sum_{\mu,\nu \in s,p,d} \sum_{i \in \text{Ga}, j \in \text{O}} \text{COHP}_{\mu i, \nu j}(\varepsilon) \qquad (1)$$

$$\text{COHP}_{\text{O}-\text{O}}(\varepsilon) = \sum_{\mu,\nu \in s,p,d} \sum_{i \neq j \in \text{O}} \text{COHP}_{\mu i, \nu j}(\varepsilon) \qquad (2)$$

notice that in Eq. (2) we avoid the self-interaction ($i \neq j$). Fig. 1(e) shows these two COHPs, both calculated from DFT wavefunctions. It can be seen that the VBM is dominated by an anion-anion antibonding coupling (AAAC) of O-O rather than the conventional Ga-O bonding coupling, the latter of which rises only when going into deeper energy of valence bands. The existence of AAAC can be illustrated by another way, that is to calculate the decomposition of the total COHP on wave vectors **k**, i.e., $\text{COHP}(\varepsilon, \mathbf{k})$. It allows us to overlay the COHP onto the electronic band structure to investigate the distribution of orbital coupling in momentum space. It can be seen from Fig. 2(a) that the COHP of VBM show significant antibonding characteristics along the whole BZ. The most negative value is -0.182 on $\Gamma$, while on $I$ (global VBM) it is -0.08. As we delve deeper into the matrices of $\text{COHP}_{\mu i, \nu j}(\text{VBM}, \Gamma)$ and $\text{COHP}_{\mu i, \nu j}(\text{VBM}, I)$ (Fig. 2(b) and Fig. 2(c)), we find that such antibonding coupling on VBM come from very specific oxygen pairs. For VBM on $\Gamma$, the antibonding coupling is primarily characterized by the O-O $p_z$-$p_z$ interactions (the top-right triangle region in Fig. 2(b)) among which the strongest one is between $O_2$-$O_5$ pair, while all other couplings (the Ga-O and Ga-Ga regions) are negligible. While for VBM on $I$ the dominant contribution is from the $p_x$-$p_x$ antibonding between $O_1$-$O_4$.

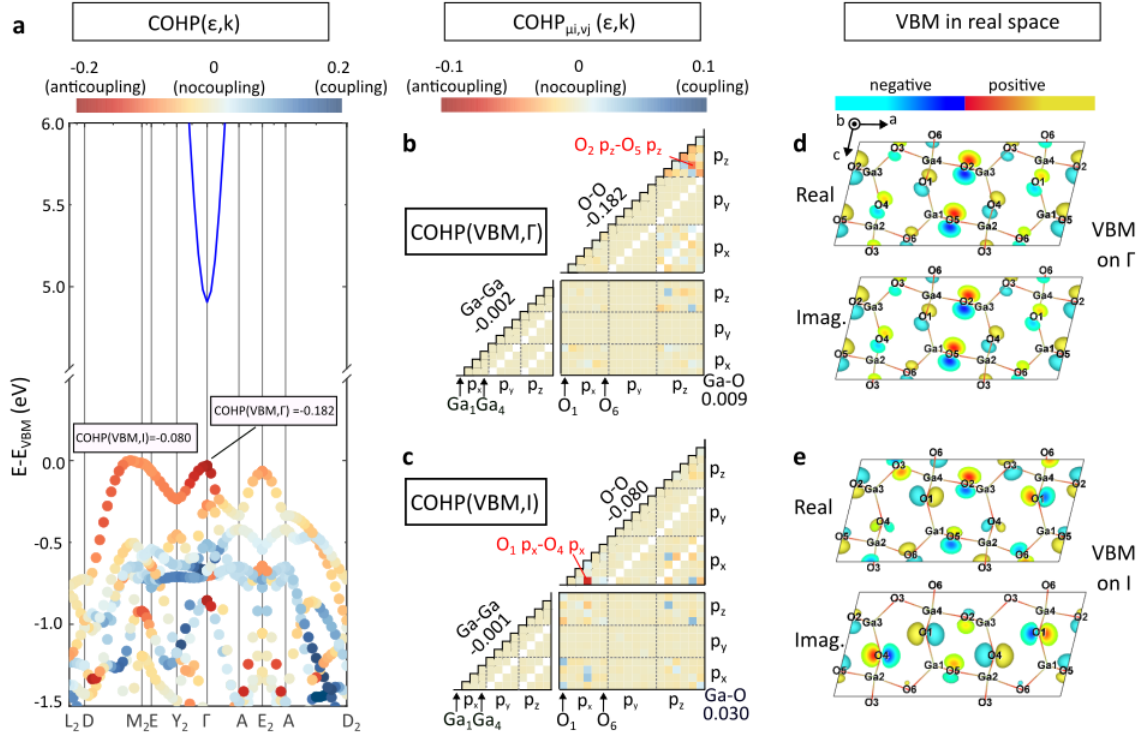

**FIG. 2. The antibonding coupling between different oxygens**. (a) The $\text{COHP}(\varepsilon, k)$ plotted on band structure showing the density of COHP as the function of energy and wave vector. The two numbers in small frames are the $\text{COHP}(\varepsilon, k)$ value of VBM at $I$ and $\Gamma$, respectively. The overall $\text{COHP}(\varepsilon, k)$ can be furtherly decomposed into matrix $\text{COHP}_{\mu i, \nu j}(\varepsilon, k)$ for different orbital pairs $\mu i\text{-}\nu j$, as visualized by triangle matrices in (b) for VBM at $\Gamma$, and in (c) for VBM at $I$, where each solid square represents one coupling, going through all Ga-p and O-p orbitals. Note that all diagonal squares are blank as the self-interactions are ignored. Among all the coupling, the $O_2\text{-}O_5$ $p_z\text{-}p_z$ and the $O_1\text{-}O_4$ $p_x\text{-}p_x$ coupling are the strongest ones and hence marked by red texts. The three numbers in (b) (as well as in (c)) are the sum of the matrix elements over all Ga-Ga coupling, all Ga-O coupling, and all O-O coupling, respectively. The distributions in real space of the real and imaginary parts of VBM wavefunction at $\Gamma$ (d) and $I$ (e), with all Ga and O atoms labeled.

Such an AAAC effect in β-Ga$_2$O$_3$ seems to be non-trivial, given that even the closest oxygens are separated by over 3 Å, a distance much larger than their covalent radius 0.7 Å. In fact, it can be understood by the following two facts.

*(a) The unique crystallographic feature of Ga$_2$O$_3$.* Fig. 2(d) and Fig. 2(e) show the atomistic model of the conventional cell of β-Ga$_2$O$_3$, together with the real-space distribution of VBM wavefunctions on $\Gamma$ and on $I$, respectively. It can be seen that between the O$_2$ and O$_5$ - where the strongest AAAC occurs - there are "void" region along the c axis, that these two oxygens are not only close to each other (3.0 Å, which is the shortest distance among all O-O pairs), but have no cations (Ga) in between them, leaving enough space for their $p_z$ orbitals to interact with each other; the same situation exists for $O_1$ and $O_4$ along the $a$ axis.

*(b) The symmetry and ligand orbital interactions.* Firstly, the β-$Ga_2O_3$ has space group $C2/m$ with severely distorted octahedral and tetrahedral Ga-O local clusters, and hence relatively weak symmetry restrictions on orbital coupling, i.e., almost all O and Ga orbitals can interact with each other. Secondly, for each formula unit, the two Ga atoms have in total 8 orbitals that can be paired (one *4s* and three *4p* for each Ga, while *3d* are too deep in energy to be involved, see supporting information Fig. S2), while the three O atoms have in total 9 orbitals (three *2p* for each O, while *2s* are too deep in energy). It means that there always leaves one O orbital not coupled with Ga orbitals per formula unit; these "leftovers" then could form O-O bonding/antibonding states if they meet the proper conditions with respect to symmetry and distance.

## IV. THE AAAC ENHANCES THE HOLE MASS AND THE FORMATION OF STH.

Now, given these findings in $Ga_2O_3$ as discussed above, at first glance, one may think that *(i) the AAAC effect* is a consequence of *(ii) the anion-dominated VBM*, and so as *(iii) the heavy hole mass*, i.e., *(ii)* results in *(i)* and *(iii)*. However, such causality is in fact reverse, that *(i)* actually results in *(ii)* and *(iii)*. We provide the discussion as below:

*(1) AAAC effect results in the oxygen-dominated VBM.* Indeed, an anion-dominated VBM can be found in other compounds, e.g., ionic compounds and ligand-hole transition-metal oxides ( [43]) such as titanates, vanadates, and nickelates. $Ga_2O_3$, however, does not belong to either case. Firstly, the strong Ga-O bonding effect on deeper valence bands (Fig. 1(e), where the Ga-O COHP is positive) shows a covalent nature rather than ionic. Secondly, in a ligand-hole transition-metal oxide, the symmetries of crystal allow bonding/antibonding between *O-p* and transition metal *d*-orbitals that have the same irreducible representation, but forbid those who do not; these bonding *O-p* and metal-*d* orbitals are hence pushed by Coulomb repulsion into deeper/higher energies, and leave a VBM holding no-bonding O-*p* orbitals. The low-symmetric $Ga_2O_3$, as we discussed above, cannot host such strong symmetry restrictions on orbital coupling. Therefore, if one only considers the metal-oxygen coupling as in the conventional case, it should not have strong enough Coulomb repulsion to separate in energy the O and Ga contributions and make an oxygen-dominated VBM. However, the existence of AAAC effect now offers extra repulsion from the O-O bonding states below, which eventually pushes the AAAC state up in energy to form the VBM.

(2) *AAAC effect leads to hole mass enhancement.* After the AAAC VBM being pushed up in energy, the internal gap between the VBM and other valence bands are opened by the avoid-crossing mechanism. As a consequence, the VBM state has a narrow bandwidth <1 eV (as shown in Fig. 2(a)). This could introduce a significant mass enhancement for the holes. A brief intuition originates from that the factors that control effective masses can be gleaned qualitatively from the $\mathbf{k} \cdot \mathbf{p}$ perturbation theory [44],

$$\frac{1}{m^*_{\text{VBM},\mathbf{k}\alpha}} = \frac{1}{m_0} + \frac{2}{m_0^2} \sum_{n \neq \text{VBM}} \frac{\left|\langle \text{VBM}, \mathbf{k} | p_{\|\alpha} | n\mathbf{k} \rangle\right|^2}{\varepsilon_{\text{VBM},\mathbf{k}} - \varepsilon_{n\mathbf{k}}} \tag{3}$$

where $m^*_{\text{VBM},\mathbf{k}\alpha}$ is the effective mass of VBM at wave vector $\mathbf{k}$ along direction $\alpha$, $m_0$ is the bare electron mass, $\langle \text{VBM}, \mathbf{k} | p_{\|\alpha} | n\mathbf{k} \rangle$ is the transition dipole moment (TDM) between VBM and state $n$ along direction $\alpha$, and $\varepsilon_{\text{VBM},\mathbf{k}}$ and $\varepsilon_{n\mathbf{k}}$ are the eigenvalue of VBM and state $n$, respectively. The sum is over all states except VBM itself. A trivial case in semiconductors is when there exists a large and negative term at the Van Hove singularity between CBM-VBM due to the strong TDM while all other terms are relatively weak. The sum is hence significantly negative and eventually results in a light and negative $m^*_{\text{VBM},\mathbf{k}\alpha}$, *i.e.*, a light hole state. While in β-Ga$_2$O$_3$, (i) the wide gap enhances the denominator $\varepsilon_{\text{VBM},\mathbf{k}} - \varepsilon_{\text{CBM},\mathbf{k}}$, while (ii) the AAAC effect weakens the numerator between CBM-VBM and induces extra positive terms between VBM and deeper valence bands. Here we would take the VBM at $\Gamma$ point as an example. Note that the space group of β-Ga$_2$O$_3$ is $C2/m$, while the little group at $\Gamma$ is $C_{2h}$ and has 4 irreducible representations $A_g$, $A_u$, $B_g$ and $B_u$. As discussed above and shown in Fig. 2(d), in the real space the VBM at $\Gamma$ mainly distributes between O-O pairs distanced from Ga nuclei, which suppresses the overlap in real space between VBM and CBM (the latter being Ga *4s* orbitals). Moreover, Fig. 2(d) reveals that the VBM at $\Gamma$ has an irreducible representation of $B_u$ ($\Gamma_2^-$). The symmetry analysis then teaches that states that can hold non-zero TDM to VBM would be $A_g$ ($\Gamma_1^+$) and $B_g$ ($\Gamma_2^+$), while within the range of 2 eV below VBM there are many O-$p$ VBs with these two representations (Table SI), leading to extra positive terms in the sum in Eq. (3), and compensating the negative term from CBM-VBM coupling. In the end, the heavy hole mass at $\Gamma$ can be a consequence of the sum in Eq. (3) being only slightly negative. This approach can also explain the positive hole mass (inverted sign) at k-points such as $M_2$, $A$ and $Y_2$, as at these k-points the CBM-VBM gap goes larger while the VBM-deeper-VB gaps are relatively consistent, in which case the sum in Eq. (3) becomes positive and, eventually, $m^*_{\text{VBM},\mathbf{k}\alpha}$ becomes positive.

*(3) AAAC effect enhances the formation of STH.* The acceptors in Ga$_2$O$_3$, once ionized, will contribute excess holes to occupy firstly the AAAC VBM. The O-O antibonding strength will hence be weakened, and the distance between these O-O pairs are shortened. As the AAAC VBM is highly localized, it will induce local distortions, which will in-return trap the excess holes. Such a positive feedback will eventually enhance the formation of STH.

## V. THE DESIGN PRINCIPLE TO TUNE THE HOLE MASS IN β-Ga₂O₃

In previous section, we have investigated the origin of AAAC effect and its consequence on the hole mass enhancement and formation of STH. It then enlightens us the design principle if one could use the AAAC effect as a "route to light holes" in β-Ga$_2$O$_3$. As can be seen from Fig. 2(a)(b), the VBM at $\Gamma$ point has O-O $p_z$-$p_z$ AAAC in the majority and hosts a much lighter hole mass, but is slightly lower in energy than the global VBM on $I$, the latter of which has the heavy mass and AAAC between O-O $p_x$-$p_x$. Therefore, by (1) enhancing the O-O $p_z$-$p_z$ coupling and/or (2) suppressing the $p_x$-$p_x$ ones, one can reverse the order in the energy of the VBM at $\Gamma$ point and the one on $I$, and hence achieves a light hole mass. Following this design principle, one of the simplest ways to control the O-O coupling is to use strain to tune the O-O distance along different directions. Therefore, in the following sections, we show the predictions on hole mass and STH formation under strains.

## VI. STRAIN-INDUCED LIGHT HOLES AND THE QUASI-TWO-DIMENSIONAL TRANSPORT BEHAVIOR

We have tested all types of uniaxial, biaxial, and hydrostatic strains, and found that the most efficient strain to lighten to hole mass is (a) uniaxial tensile strain along the $b$ axis and (b) biaxial compressive strain on $a$-$c$ plane. As the evolutions of hole properties are almost identical under (a) and (b) but differ only by the value of the critical strain, in the following sections we will discuss (a) as the example. Cases for other strains are summarized in the supporting information. As illustrated in Fig. 3(a), with the tensile strain increasing, the band gap decreases, while the VBM on $\Gamma$ rises in energy faster than that on $I$ point. At a critical strain of 1.5%, the VBM on $\Gamma$ becomes the new global VBM, which also indicates a transition from indirect to direct band gap, as demonstrated in Fig. 3(c). Such band edge transition is also observed in the compressive biaxial strain on the $a$-$c$ plane with a smaller critical strain of 0.7% (see Fig. S3 (a)(b)). As the consequence of the shift of VBM in momentum space, we observe a direct, step-function-like changes in $m_h^*$ near the critical strain (Fig. 3(b)). For instance, the hole mass along $c^*$ ($m_{h\parallel c^*}^*$) drops suddenly from 4.77 to 0.38 $m_0$, decreasing to 8% of its original value brings it comparable to the effective mass of electrons ($m_e^*$ = 0.28 $m_0$). This further causes the conductivity effective mass ($m_h^*$) decreasing from the 3.47 $m_0$ to 0.99 $m_0$.

Furthermore, we also find that the VBM now becomes highly anisotropic, e.g., the hole mass along $b^*$ is approximately 70 times the mass along $c^*$, $m_{h\parallel b^*}^* \approx 70 m_{h\parallel c^*}^*$. It potentially reveals a quasi-two-dimensional transport behavior within the 3-dimensional bulk Ga$_2$O$_3$. The low-dimensional behavior of holes can also be visualized from the Fermi surfaces shown in Fig. 3(d), (e). In both the unstrained and the 2%-tensile-strained cases, the Fermi surfaces are chosen as the isosurfaces of valence bands at -0.05 eV (~2kT at 300 K) below VBM to mimic the p-doped crystal at room temperature. Note that the more spreading the hole has in momentum space, the more localization and difficulty in transport it has in real space. The unstrained Fermi surfaces spread almost isotropically in the whole first BZ (i.e., heavily localized in all directions in real space), while the tensile-strained Fermi surfaces show much smaller spreading along $a^*$ and $c^*$, indicating the hole mass lightening along these two directions.

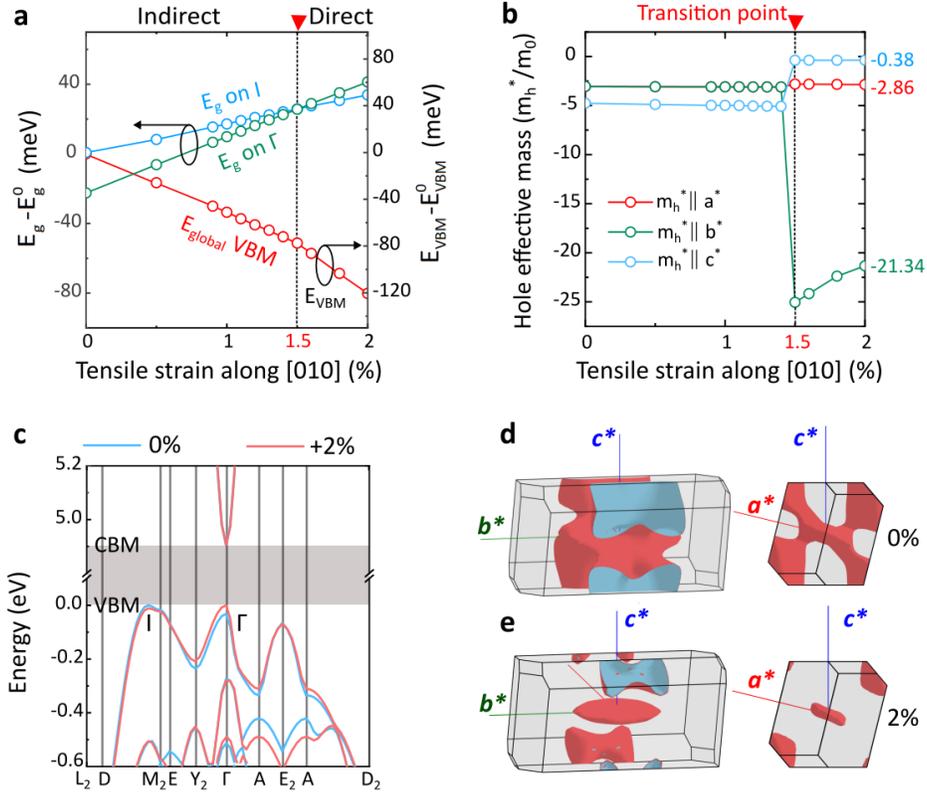

**FIG. 3. Strain-induced modulation on VBM in β-Ga$_2$O$_3$.** (a) Changes in band gap and VBM energies at $\Gamma$ and $I$ under uniaxial strain along the $b$ axis. $E_{VBM}^0$ is the VBM energy while the superscript '0' means no strain. The vertical dash line indicates the critical strain of direct-indirect gap transition. (b) Hole masses of VBM under strain. (c) Band structure under a 2.0% tensile strain along $b$, compared to the one in the unstrained case; valence bands and conduction bands have been aligned to CBM and VBM, respectively. Fermi surfaces upon (d) zero strain and (e) 2.0% tensile strain along the $b$ axis, drawn for an energy -0.05 eV below VBM, which mimics the distribution of holes in a p-doped crystal.

Furthermore, we find that the evolutions of $m_h^*$ under strain indeed follow our design principle, that they are the consequence of the modulation on AAAC strengths. As the strongest AAAC is on O$_2$-O$_5$, followed by the ones on O$_1$-O$_4$, O$_1$-O$_6$, and O$_3$-O$_4$ (see the labels for all Ga and O in Fig. 4(a)), we trace under strain the variation of distance between these O pairs (Fig. 4(b)), and the differential of COHP ($\Delta$COHP; we use the COHP under zero strain as the reference) on them (Fig. 4(c)-(e)). It can be seen from Fig. 4(b) that when applying the uniaxial tensile strain along the $b$ axis, the distance between O$_1$-O$_4$ is elongated, while the ones between O$_1$-O$_6$ and O$_3$-O$_4$ are shortened. The distance between O$_2$-O$_5$ only shows negligible changes under all strains calculated. As the consequence shown in Fig. 4(d), (e), the AAAC strength on O$_1$-O$_4$ is weakened by such strain, with the AAAC on O$_1$-O$_6$ and O$_3$-O$_4$ strengthened, and the one on $O_2$-$O_5$ almost unchanged. All other COHP matrix elements shows negligible variations. It results in the increase of the overall AAAC on VBM at $\Gamma$, and the decrease of such property on VBM at $I$, as shown in Fig. 4(c). Eventually, under the tensile strain along the $b$ axis, the VBM at $\Gamma$ has been pushed up in energy in an amplitude larger than VBM at $I$, making it the new global VBM after the critical strain of 1.5%, and leading to the step-function change on hole mass, all in agreement with the band structure calculations shown in Fig. 3.

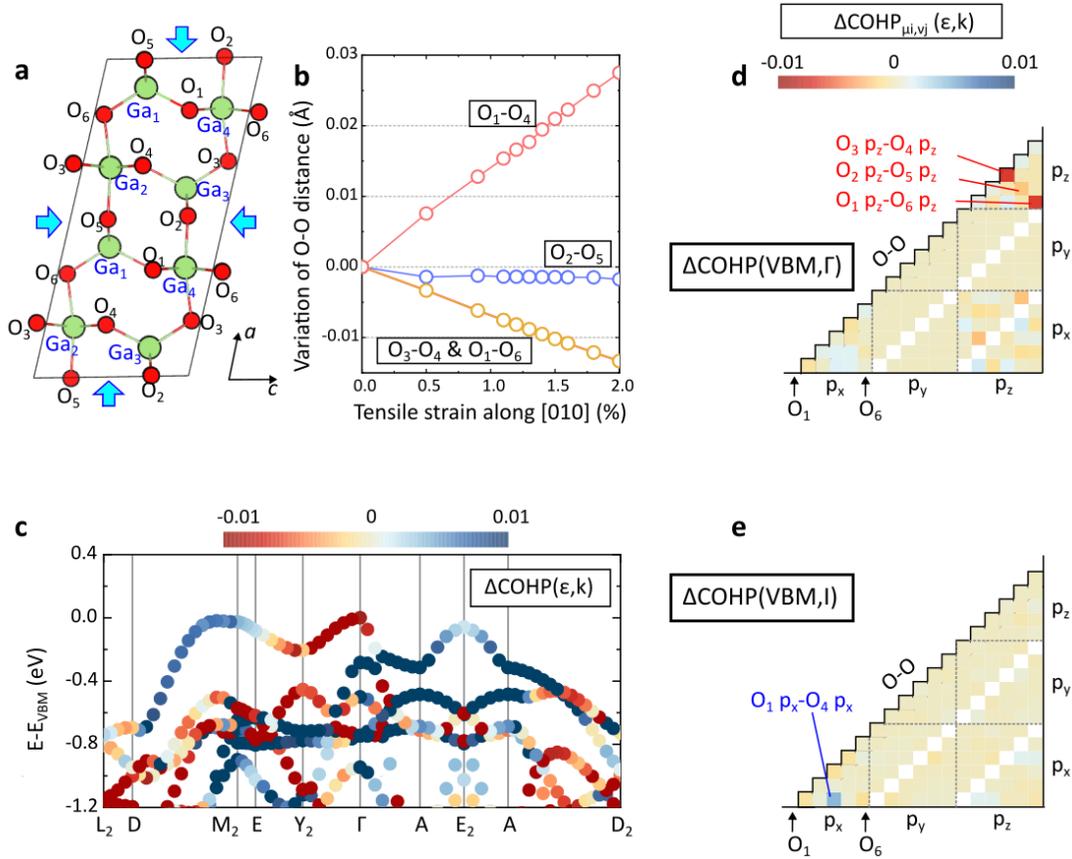

**FIG. 4. Strain-induced modulation on lattice structure and COHP in β-$Ga_2O_3$.** (a) The side view of the conventional cell, where all Ga and O are labeled. (b) Variations of distance between oxygen pairs $O_2$-$O_5$, $O_1$-$O_4$, $O_1$-$O_6$, and $O_3$-$O_4$; these oxygen pairs are the ones that have strong AAAC. (c) The differential $\text{COHP}(\varepsilon,\mathbf{k})$ ($\Delta\text{COHP}(\varepsilon,\mathbf{k})$) under strain. The values under zero strain have been taken as the references. The matrix elements of $\Delta\text{COHP}(\varepsilon,\mathbf{k})$ for VBM at $\Gamma$ (d) and VBM at $I$ (e).

## VII. THE SUPPRESSION OF STH

Given that strain conditions can weaken the AAAC effect, resulting in a step-function-like reduction in the hole effective mass of β-$Ga_2O_3$, we then investigate the strain effect on hole polarons in β-$Ga_2O_3$.

Indeed, the formation energy of polaron is known to be related to the effective mass $m^*$ of the host state. For instance, it has been suggested by D. W. Davies *et al.* [45] and W. H. Sio *et al.* [46] that $E_{\text{polaron}} \propto -m^*$, that the heavier the mass is on one band, the easier the polaron forms from there. Nevertheless, the formation of STH in wide-gap materials could be more complex, and simple approach could be incorrect or imprecise to describe it. Here, to understand if the tailoring of AAAC and $m_h^*$ can also affect the STH, we calculate before and after strain the formation energy of hole polaron from DFT with HSE. Details are given in the Appendix section.

Fig. 5(a) presents the hole polaron in the unstrained, pristine β-Ga$_2$O$_3$ with no vacancy or impurity. The STH is heavily localized between O$_2$ and O$_5$. Notice that it is the oxygen pair that has the strongest AAAC. It then supports our conclusion that a strong AAAC will lead to STH. Our results predict a 0.57 eV formation energy ($\varepsilon_{\text{STH}}$) for such STH, with the two oxygens distorted from their equilibrium positions by 0.22 Å. With a 2% tensile strain along the $b$ axis, the STH still exists, but its formation energy has been suppressed by 0.12 eV. The local distortions are also smaller than those under 0 strain. Moreover, the formation energy shows a relatively significant drop between 1% and 2%, which is also consistent with the evolution of hole mass predicted in the last section, as shown in Fig. 5(b). Although the reduction in hole polaron formation energy facilitates a slight increase in free hole density (see Section VIII), the formation energy remains significantly higher than the thermal kinetic energy at room temperature (0.026 eV). As a result, hole polarons under strain remain stable.

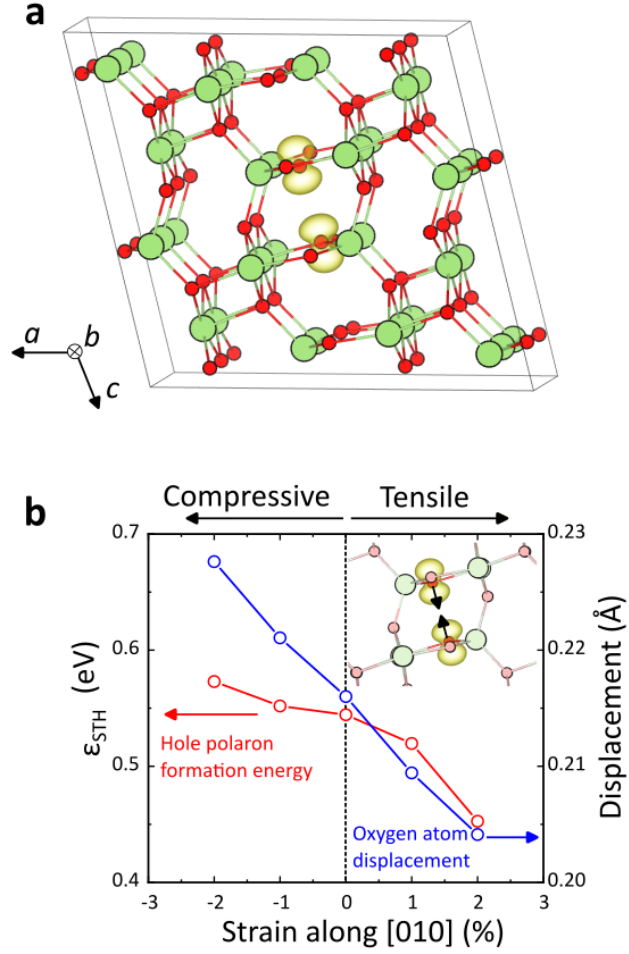

**FIG. 5. Strain-induced modulation on STH**: (a) The 80-atom supercell of β-Ga$_2$O$_3$ under 0% strain, showing the spontaneous formation of STH with the yellow isosurfaces. The isosurfaces are 0.0075 e/Å$^3$. (b) The DFT-predicted formation energy (red) and local distortion (blue) of STH under different strain along the  *b*  axis. The inset of (b) shows the local atom displacement caused by the STH, while the blue arrows mark the directions of displacement.

### VIII. The enhancement on p-type conductivity

In a semiconducting system with in-gap acceptor dopant and STH state, the density of free holes can be solved by the system of equations derived from the charge neutrality condition:

$$N_V \exp\left(-\frac{E_F - E_V}{k_0 T}\right) + \frac{N_{STH}}{1 + g\exp\left(-\frac{E_{STH} - E_F}{k_0 T}\right)} = N_C \exp\left(-\frac{E_C - E_F}{k_0 T}\right) + \frac{N_A}{1 + g\exp\left(-\frac{E_F - E_A}{k_0 T}\right)} \quad (4)$$

$$p_0 = N_V \exp\left(-\frac{E_F - E_V}{k_0 T}\right) \quad (5)$$

$$N_C = 2\left(\frac{m_e^* k_0 T}{2\pi\hbar^2}\right)^{\frac{3}{2}}, \quad N_V = 2\left(\frac{m_h^* k_0 T}{2\pi\hbar^2}\right)^{\frac{3}{2}} \tag{6}$$

where $p_0$ is the free hole density in valence band, $E_F$ is the fermi level, $E_V$ and $E_C$ are the energies of VBM and CBM, respectively, $E_A$ is the acceptor level, $E_{STH}$ is STH level, $k_0$ is the Boltzmann constant, $T$ is the temperature, $g$ is the degeneracy factor, $N_A$ is the density of acceptor dopants, $N_V$ and $N_C$ are the effective densities of states in VB and CB, respectively, and $m_e^*$ and $m_h^*$ are the effective masses of CBM and VBM, respectively. Considering the physical limitation that the density of STH cannot exceed the total density of ionized holes, the system of Eq. (4)-(6) can be solved numerically.

For this analysis, we use parameters from Mg-doped β-Ga$_2$O$_3$ ($E_A = 1.05$~$1.25$ eV [47], $N_A = 10^{18}$~$10^{19}$ cm$^{-3}$), and results from this work ($m_h^* = 3.47 m_0$ and $E_{STH} = 0.57$ eV before strain; $m_h^* = 0.99 m_0$ and $E_{STH} = 0.42$ eV after strain; $m_e^* = 0.28 m_0$; $N_{STH} \approx 10^{21}$ cm$^{-3}$; band gap is 4.9 eV; $g = 2$ in the low-symmetric lattice; and $T = 300K$, room temperature). We find that the density of free holes $p_0$ increases by approximately 55% after strain, primarily due to the 0.12 eV reduction in STH formation energy, while the decrease in hole mass $m_h^*$ contributes negligibly to the variation of $p_0$.

The p-type conductivity is given by:

$$\sigma_p = p_0 q \mu_p, \quad \mu_p = \frac{q\tau_p}{m_h^*} \tag{7}$$

where $q$ is the elementary charge, $\mu_p$ is the hole mobility, and $\tau_p$ is the relaxation time of free holes. Assuming that $\tau_p$ is independent of strain conditions, it follows that $\sigma_p \propto \frac{p_0}{m_h^*}$. Finally, our analysis reveals that the p-type conductivity $\sigma_p$ exhibits an increase of approximately 443% under strain, mainly due to the reduction of hole mass from 3.47 m$_0$ to 0.99 m$_0$. Notably, this enhancement is achieved without altering the acceptor level $E_A$, which, as a deep level acceptor state, still imposes a significant limitation on the overall p-type conductivity.

## IX. SUMMARY AND OUTLOOK

In this work, we identify the physical origin of the intrinsic, heavy hole mass in β-Ga$_2$O$_3$ as an anion-anion antibonding coupling (AAAC) among oxygen atoms. This coupling, occurring at distances longer than atomic covalent radii, arises from the low symmetry and ligand orbital interactions, where oxygen orbitals interact without gallium involvement. The resulting Coulomb repulsion pushes the antibonding states up in energy to form the VBM, narrowing its bandwidth to less than 1 eV, significantly enhancing the hole mass and leading to STH.

Based on these insights, we propose a design principle to tune hole masses and STH by modifying the strength of AAAC. We find that specific strain conditions can reduce the hole mass along $c^*$ from 4.77 $m_0$ to 0.38 $m_0$, tailoring its mass to a value comparable to the electron mass in the same compound. This significant reduction of hole mass along $c^*$ leads to the conductivity mass decreases from 3.47 $m_0$ to 0.99 $m_0$. The critical strains are 1.5% for uniaxial tensile strain along *the b axis* and 0.7% for biaxial compressive strain on $a$-$c$ plane. This strain modification also induces high anisotropy in the hole mass, enabling quasi-two-dimensional transport in bulk material. Furthermore, strain slightly reduces the formation energy of hole polarons from -0.57 eV to -0.45 eV, contributing to an increase in free hole density. However, the relatively high formation energy of hole polarons under strain suggests that they remain stable.

These findings provide a new perspective on controlling hole transport properties in novel wide-gap materials, especially those with low-symmetric crystallographic structures, potentially leading to more efficient electronic and optoelectronic devices. We expect future research to validate these findings in gallium oxide and explore other material systems where similar mechanisms might be exploited to enhance performance.

## ACKNOWLEDGEMENTS

This work was supported by the National Key R&D Program of China (2022YFB3605400), the National Natural Science Foundation of China (62234007, 62293521, U21A20503, U21A2071 and 12174380). We would like to thank Lin-Wang Wang from Institute of Semiconductors, Chinese Academy of Sciences for insightful discussions.

## COMPETING INTERESTS

The authors declare no competing interests.

## APPENDIX I: Computational parameters

The crystallographic structure of β-Ga$_2$O$_3$ was firstly obtained from the Materials Project database [48], then fully relaxed by ourselves. All calculations in this work were implemented within the framework of density functional theory (DFT) implemented in the Vienna ab initio simulation package (VASP) [49] using the projector augmented wave (PAW) pseudopotentials [50]. A hybrid functional within the Heyd–Scuseria–Ernzerhof (HSE06) method was used in all DFT calculations, with a mixing parameter of 0.35 for the Hartree-Fock exchange and a screening parameter of 0.20, consistent with those applied in previous reports [6,19,51]. It has been demonstrated in the discussion of this work that this parameter set can reproduce the experimental observations of both the band gap and the energy of the polaronic state (band gap: 4.91 eV from this work; 4.84 eV This parameter set reproduces the experimental observations of both the band gap and the energy of the polaronic state (band gap: 4.91 eV from this work; 4.84 eV from [52]; polaronic state: 4.34 eV below CBM from this work; 4.40 eV below CBM from [52]), indicating a fulfilling of the generalized Koopmans' theorem. A complete fulfillment of the generalized Koopmans' theorem could be achieved by fine-tuning the HSE parameters, as reported in [53]. The cutoff energy of plane-wave basis was 520 eV in all runs. An 8×8×4 Monkhorst-Pack k-point mesh was used for conventional cell. Energy minimization and ionic relaxation were performed with a tolerance of $10^{-8}$ eV per formula unit for the total energy, and $10^{-6}$ eV/Å on each nuclear for the atomic force, respectively. The shifts in energy of eigen levels in Fig. 3(a) under different strains were calculated by aligning the electrostatic potential (core level). The effective mass $m^*$ was calculated by fitting a quadratic curvature at the band extrema $(m^*)^{-1} = \partial^2 E/\partial k^2/\hbar^2$.

**APPENDIX II: Basic concept of COHP**

The crystal orbital Hamilton populations (COHPs) was calculated by the Local-Orbital Basis Suite Towards Electronic-Structure Reconstruction package (LOBSTER) [54,55]. Here, we offer a brief explanation of the COHP method [56]:

$$\text{COHP}_{\mu i, \nu j}(\varepsilon, \mathbf{k}) = \sum_l \mathcal{R}\left[P^{(proj)}_{\mu i, \nu j, l}(\mathbf{k}) H^{(proj)}_{\nu j, \mu i}(\mathbf{k})\right] \times \delta(\varepsilon_l(\mathbf{k}) - \varepsilon) \qquad (7)$$

where $\mu$, $\nu$ are the atomic orbitals of the $i^\text{th}$ and $j^\text{th}$ atoms, respectively, $i$ is the index of electronic band, and $\varepsilon_l(\mathbf{k})$ is the eigen energy of the $i^\text{th}$ band at wave vector $\mathbf{k}$. The $H^{(proj)}_{\nu j, \mu i}$ on the right-hand side is the Hamiltonian matrix elements expressed in the basis of the local functions. The $P^{(proj)}_{\mu i, \nu j, l}$ is the projected density matrix

$$P_{\mu i,\nu j,l}^{(proj)}(\mathbf{k}) = T_{l,\mu i}^*(\mathbf{k})T_{l,\nu j}(\mathbf{k}) \qquad (8)$$

and $T_{l,\mu i}(\mathbf{k})$ is the transfer matrix

$$T_{l,\mu i}(\mathbf{k}) = \langle \varphi_l(\mathbf{k})|\phi_{\mu i}\rangle \qquad (9)$$

where $\varphi_l(\mathbf{k})$ is the eigen wavefunction of the $i^{th}$ band at wave vector $\mathbf{k}$, and $\phi_{\mu i}$ is the $\mu^{th}$ local orbitals of the $i^{th}$ atom.

**APPENDIX III: Polaron formation energy**

The hole polarons in $Ga_2O_3$ were calculated using the 80-atom supercell (the 1×2×2 extension of the conventional cell). The steps are as follows. (1) An electron was firstly removed from the supercell to mimic the hole injection while all ions were frozen at their equilibrium positions. Mark the total energy for such a frozen-ion system with an extra hole as $\varepsilon_{tot}(\{\mathbf{R}_0\};\eta^+)$. (2) Secondly, random displacements were applied on ions as the initial "kicks". (3) Internal atomic positions were then relaxed with fixed lattice vectors. Mark the new total energy as $\varepsilon_{tot}(\{\mathbf{R}'\};h^+)$. During all steps the spin polarization was considered. The formation energy of polaron can be calculated by

$$\varepsilon_{STH} = \varepsilon_{tot}(\{\mathbf{R}_0\};\eta^+) - \varepsilon_{tot}(\{\mathbf{R}'\};h^+) + \langle\Delta V\rangle \qquad (10)$$

where

$$\langle\Delta V\rangle = \langle V(\{\mathbf{R}_0\};\eta^+)\rangle - \langle V(\{\mathbf{R}'\};h^+)\rangle \qquad (11)$$

are the difference in the average electric potential of the two systems [6].

We have also calculated the formation energy using the method proposed by Christoph Freysoldt et al [57], which considers the charged-defect correction using both the long-range electrostatic interactions induced by periodic boundary conditions, as well as the short-range potential effects within the supercell caused by the charged defect. We have found that both methods give almost identical results.